\documentclass{kapproc} 

\usepackage{procps} 

\usepackage[dvips]{graphicx}

\upperandlowercase

\normallatexbib 

\begin{document}

\articletitle{Interband scattering in M\MakeLowercase{g}B$_2$}


\author{R.P.S.M. Lobo,$^1$ M. Elsen,$^1$ P. Monod,$^1$ 
J.J. Tu,$^2$ Eun-Mi Choi,$^3$ \\
Hyeong-Jin Kim,$^3$ W. N. Kang,$^3$ Sung-Ik Lee,$^3$
R.J. Cava,$^4$
and G.L. Carr$^5$}
\affil{$^1$Laboratoire de Physique du Solide (UPR5 CNRS) ESPCI, 
75231 Paris cedex 05, France \\
$^2$Department of Physics, The City College of New York, New York, NY 10031\\
$^3$National Creative Research Initiative Center for Superconductivity, 
Department of Physics, Pohang University of Science and Technology, 
Pohang 790-784, Korea\\
$^4$Princeton University, Princeton, NJ 08544\\
$^5$National Synchrotron Light Source, 
Brookhaven National Laboratory, Upton, NY 11973
}
\email{email: lobo@espci.fr}

\begin{abstract}
The scattering process responsible for connecting the bands
remains one of the last open questions on the physical properties
of MgB$_2$. Through the analysis of the equilibrium and photo-induced
far-infrared properties as well as electron spin resonance of 
MgB$_2$ we propose a phonon mediated energy transfer process between 
the bands based on the coupling of quasiparticles to an $E_{2g}$ phonon.
\end{abstract}


\section{Introduction}
The discovery of superconductivity in MgB$_2$ \cite{Akimitsu} triggered
a major theoretical and experimental effort to understand its mechanisms.
Kortus {\it et al.} \cite{Kortus} predicted that that MgB$_2$ should have
two metallic bands at the Fermi energy, one two dimensional ($\sigma$) 
band along the $\Gamma$--$A$ direction and another three dimensional ($\pi$)
band covering most of the Brillouin zone volume. They also predicted that 
both bands should undergo the superconducting transition with different
energy gaps, the higher being in the $\sigma$ band. Several experimental 
studies have found evidence for a double gap in MgB$_2$ \cite{STM, MultiGap}. 
In particular, Giubileo
{\it et al.} STM data \cite{STM} showed that these two gaps close at the same
$T_c$. The first attempt to describe a two band superconductor was done
as early as 1959 \cite{Suhl}, when Suhl {\it et al.} proposed that when
two superconducting bands interact the critical temperature ($T_c$) of the 
lowest gap band increases and joins the one of the higest gap band. This is 
qualitatively the effect observed in STM measurements. Most of the physics 
behind superconductivity of MgB$_2$ is now understood \cite{Choi} and the 
remaining big question concerns the process responsible for connecting the
two bands.

In this work we address this issue by looking at the far-infrared 
equilibrium and photo-induced reflectivity of a MgB$_2$ thin film and 
at the Electron Spin Resonance (ESR) of powder samples.

\section{Experimental}

Optical data was taken on a $\sim 30$ nm thick c-axis oriented film of 
MgB$_2$ deposited on a sapphire substrate at Postech \cite{Kang}. Its $T_c$ of 30 K 
is suppressed compared to the bulk material ($T_c$ = 39 K). Standard 
reflectance measurements were performed using the Bruker 66v FTIR 
spectrometer at beamline U10A of the NSLS, with synchrotron radiation 
as the IR source. The specimen was solidly clamped with indium gaskets 
to the copper cold-finger of a heli-tran cryostat, leaving a 3 mm 
diameter aperture exposed for the IR measurement. The remaining 80\% 
of the sample's surface was available for thermal conduction into the 
cold finger. 

Transient far-infrared photo-reflectance measurements were also performed 
using the same spectrometer and beamline. Here, we measured the reflectance 
change due to illuminating the MgB$_2$ film with 2 ps near-infrared 
($\lambda = 760$ nm) pulses from a Ti:sapphire laser. The laser pulses 
break pairs and weaken the superconducting state for a brief amount of 
time ($\sim 1$ ns) \cite{Carr}. The resulting change in reflectance is 
sensed with the $\sim 1$ ns infrared pulses from the synchrotron in a 
pump-probe configuration \cite{Lobo}. The average laser power was 20 to 
50 mW (0.4 to 1 nJ per pulse) so that the MgB$_2$ film could be maintained 
at $T = 5$ K. The pulse repetition frequency was 53 MHz, matching the 
pulsed IR output from the synchrotron. For time-dependent studies, the 
broadband probe pulses were not spectrally resolved, and the response 
is an average of the reflectance across the 10 to 100 cm$^{-1}$ spectral 
range. 

ESR data was collected between 40 K and 600 K with a Bruker ESP300E 10~GHz
spectrometer equiped with an Oxford ESR He cryostat on MgB$_2$ powder samples
synthesized in Princeton. The powder was mechanically ground to avoid skin 
depth problems.

\section{Results and Discussion} 

\begin{figure}
  \begin{center}
    \includegraphics[width=12cm]{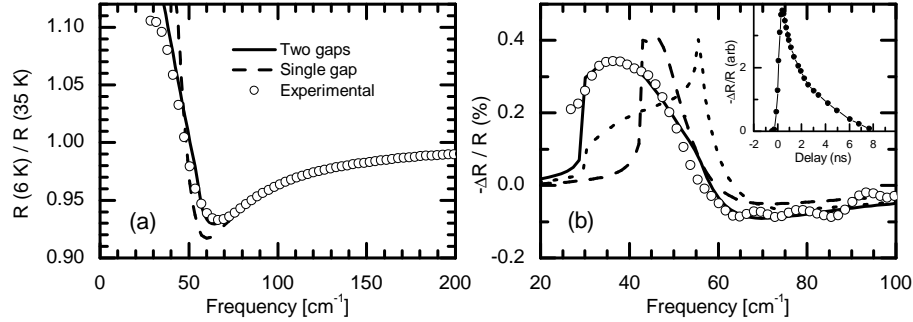} 
  \end{center}
\caption{(a) Equilibrium infrared reflectivity of MgB$_2$ film at 5 K (dots). 
The lines are fits using the Mattis-Bardeen optical conductivity for systems
with one or two gaps. (b) Photo-induced reflectivity of MgB$_2$ film at 5 K
(dots). The lines are attempts to describe the data using the same parameters
as those from panel (a). The dashed line assumes a single gap. The dotted and
continuous lines are calculations for a two gap system. For the former we 
assume that in the photo-excited state both bands have excess quasiparticles
and for the latter, we assume that all excess quasiparticles find themselves in
the lower energy gap band ($\pi$). The inset in panel (b) shows the time 
dependence of the integrated photo-induced signal.}
\label{fig1}
\end{figure}

Figure \ref{fig1}(a) shows the ratio of superconducting to normal far-infrared
reflectance at equilibrium. The dashed line is a Mattis-Bardeen fit 
\cite{Zimmermann} assuming a gap at 43 cm$^{-1}$. The solid line is obtained
using a simple non-interacting carrier model for the conductivity  
$\sigma = (1 - f) \sigma_\sigma + f \sigma_\pi$ where $\sigma_\sigma$ and 
$\sigma_\pi$ are Mattis-Bardeen conductivities for each band and $f$ is the
fraction of the $\pi$ band that contributes to the total conductivity.
The parameters used for the latter are $2 \Delta_\sigma = 56$ cm $^{-1}$ and
$2 \Delta_\pi = 30$ cm $^{-1}$ ($f = 0.6$). Although the 2 gap model better 
describes the data at low frequencies, the improvement is not sufficient to 
conclude that two gaps must exist. In fact, previous data on equilibrium 
infrared spectra of MgB$_2$ \cite{IR} did not provide any clear cut 
picture about the presence of two bands in MgB$_2$, either. However, the data 
does not seem to follow strictly the BCS theory.

An analysis of the time dependent transient photo-reflectance 
gives us a good estimate of the effective pair recombination rate \cite{Carr}. 
The inset in Fig. \ref{fig1}(b) shows the average reflectance change as a 
function of the delay between pump and probe pulses at 6 K. In accordance 
with other BCS superconductors the effective recombination time is found to 
be a few nanoseconds. One important remark is that, within the experimental 
time resolution, no evidence of multiple decays is found. In fact,  
ultra-fast pump-probe measurements on MgB$_2$ \cite{UF1, UF2} did not find 
any evidence for a double relaxation down to the ps regime. 

Panel (b) in Fig. \ref{fig1} shows the photo-induced spectra taken with pump
and probe at coincidence. The pairs broken by the laser pulse will create an
excess quasiparticle population. Owen and Scalapino \cite{Owen} showed that
this excess quasiparticle population weakens the superconducting state.
This weakened state can be spectroscopically detected and resolved as a 
slightly reduced energy gap allowing the transient photo-reflectance data to 
be analyzed using the same expressions as those for the equilibrium reflectance. 
The dashed curve in Fig. \ref{fig1}(b) assumes that the system has a single 
gap at 43 cm$^{-1}$ which is decreased by photo-excitation. Although the 
amplitude and overall shape of the signal can be reproduced by this 
simulation, quantitative agreement is not achieved. The dotted line assumes 
that the system has the same two gaps used in the equilibrium reflectance 
fit and that in the photo-excited state both gaps shrink by an amount 
consistent with a small rise in the electronic temperature. The 
introduction of this second gap does not improve the data description and, 
actually, introduces new features absent from the data. Our third approach, 
depicted by the solid line, assumes that the system does have two gaps but 
that only the smaller energy gap shrinks in the photo-excited state. This is 
the behavior one would expect if, after pair-breaking by the laser, the excess 
quasiparticles are left primarily in the band having the smaller
energy gap.

\begin{figure}
  \begin{center}
    \includegraphics[width=12cm]{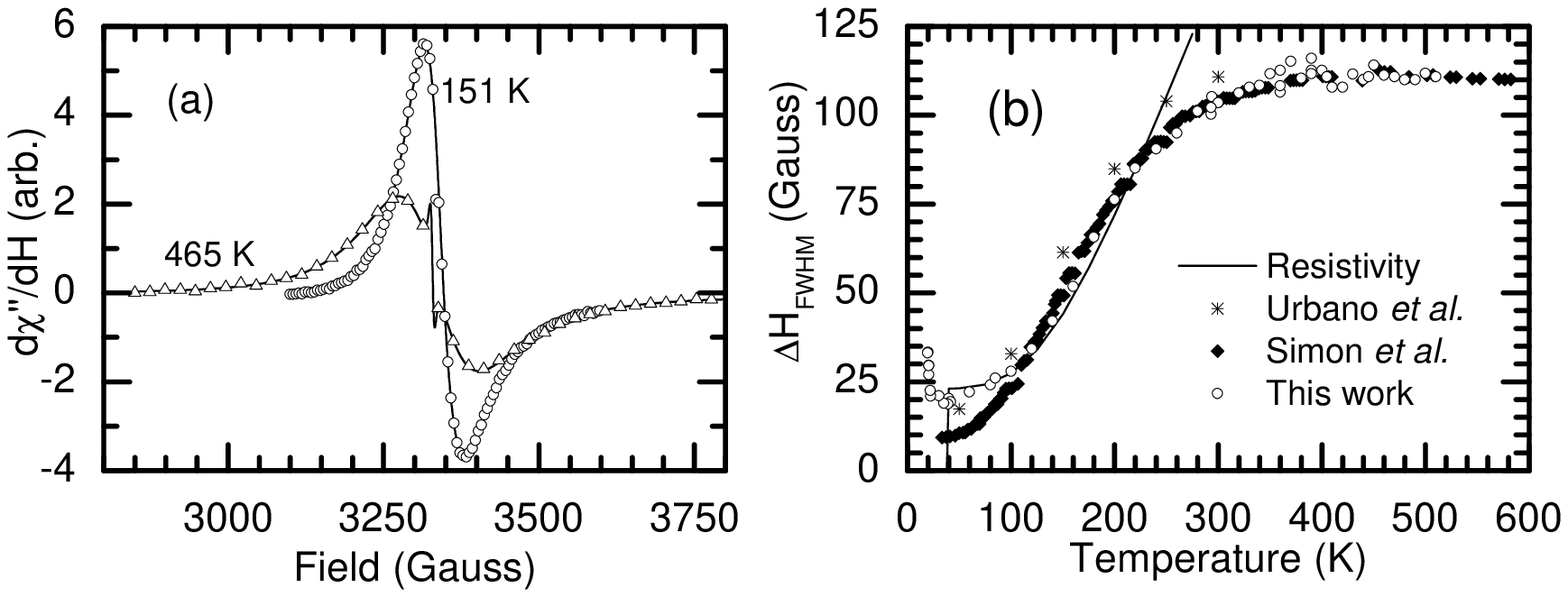} 
  \end{center}
\caption{(a) ESR signal for MgB$_2$ at 151 K (circles) and 465 K (triangles).
The solid line are fits with an assymetric Dysonian line. (b) ESR line widths
for this work (circles), Ref. \cite{Simon}. (diamonds) and Ref. \cite{Urbano}
(stars). The solid line is the resistivity (normalized to compare with the
ESR data) from Ref. \cite{Eltsev}.}
\label{fig2}
\end{figure}

Figure \ref{fig2}(a) shows the ESR signal measured for two temperatures and
fits using an assymetric Dysonian line shape. The fit allows us
to determine the Lande $g$ factor and the line width ($\Delta H$) gives us 
a measurement of the spin scattering time $\tau_s$ using 
$(g \mu_B \Delta H) \tau_s = \hbar$. 

ESR spectroscopy uses a static magnetic field to lift spin states degeneracies
by a Zeeman splitting and probes the magnetic dipole transitions between these 
states with a small perpendicular rf magnetic field. In metals, the high 
collision rate of quasi-particles defines a single spin-lattice relaxation time
$\tau_s$. The electronic spin-orbit interaction has two consequences on the ESR 
response: (i) a ``static'' interaction produces a shift in the $g$ factor from 
the free electron value and  (ii) in the presence of phonons a ``dynamic'' (time 
averaged) interaction leads to the Elliott-Yafet relation \cite{Elliott}. The 
Elliott-Yafet relation states that 
$(\delta g)^2 \tau_c / \tau_s = \textrm{const}$, where $\tau_c$ is the 
electronic collision scattering time. In a simple one band model these two 
relaxation rates are proportional, a fact that has been observed in many 
metals \cite{Monod}.

Figure \ref{fig2}(b) compares the thermal evolution of the MgB$_2$ ESR 
linewidth (this work and Refs. \cite{Simon,Urbano}) with its dc resistivity
extract from Ref. \cite{Eltsev}. The linewidth is measuring the spin scattering
rate and the resistivity gives the electron collision rate. Had Elliott-Yafet
been followed, the ESR data should agree with the resistivity. Both quantities 
agree at low temperatures but this agreement breaks down around room temperature.
Formally this implies a $g$ factor unexpectedly varying with
temperature. However, if one considers that the $\sigma$ and $\pi$ bands have
different $g$ factors, a temperature dependent mixing of these bands leads
to a temperature dependent effective $g$.

\section{Interband scattering}

The photo-induced spectra in the superconducting state and the ESR data in the
normal state can be understood in terms of an interband scattering process. When
talking about interband scattering, one may think in terms of transfer of
quasiparticles or, alternatively, transfer of the energy of quasiparticles.
The latter process can be viewed as a process in which excited quasiparticles
in one band relax emitting phonons which can excite quasiparticles in the other
band. There is no transfer of the actual quasiparticles but rather of their energy.

The recent inelastic X-ray scattering data of Shukla {\it et al.} \cite{Shukla} 
shows that the $E_{2g}$ phonon at around 60 meV is anomanously large in the $\Gamma-A$
direction of the Brillouin zone, the same direction as the $\sigma$ bands. This
$E_{2g}$ phonon is the only phonon showing a large linewidth and it is natural
to assume that this effect is coming from a strong coupling with $\sigma$ 
quasiparticles. However, once created, $E_{2g}$ phonons can scatter quasiparticles
in both bands. As the $\pi$ bands occupy a larger volume of the Brillouin zone,
it is natural to assume that more quasiparticles in this band will get excited.

This picture naturally explains our data. The photo excited spectra originates
from quasiparticles created by a 1.5 eV laser excitation. When quickly relaxing 
from high energies, the quasiparticles in the $\sigma$ band eventually reach 
energies comparable with that of the $E_{2g}$ phonon. At that time their energy 
is transfered to the $\pi$ bands leading to a photo-excited state (at our ns time
resolution) that is composed mostly of excess quasiparticles in the $\pi$ band.
The ESR response can be understood in similar terms. In this case we have no
excited quasiparticles. At lower temperatures the $g$ factor is a mixture of
$\sigma$ and $\pi$ bands. When we raise the temperature we give enough energy
to the quasiparticles to interact with the $E_{2g}$ phonon. Eventually, this
changes the balance between $\sigma$ and $\pi$ bands populations inducing a
change in the effective $g$ factor.

\section{Summary}

We looked at the far-infrared equilibrium and photo-induced reflectivity of 
a MgB$_2$ thin film and at the ESR spectra of powder samples. The 
photo-induced response shows one gap at 3 meV and another at 7 meV. However,
excess quasiparticles are only found at the edge of the smaller gap. ESR 
data on the normal state shows a break down, around room temperature, of 
the Elliott-Yafet time relaxation behavior observed in virtually any metal. 
These effects can be described in terms of an interband energy transfer process 
mediated by the anomanously broadened $E_{2g}$ phonon \cite{Shukla}.

\begin{acknowledgments}
This work was performed with the support of the U.S. Department of Energy through 
contract DE-ACO2-98CH10886 at the NSLS. Operation of the synchronized laser system 
is also supported by DOE through contract DE-FG02-02ER45984 with the University 
of Florida. We are grateful to P.B. Allen, J. Carbotte, T. Devereaux, M.V. Klein, 
I.I. Mazin, A.J. Millis, M.R. Norman, E. Nicol and D.B. Tanner, for useful 
discussions.
\end{acknowledgments}


\begin{chapthebibliography}{00}
\bibitem{Akimitsu} J. Nagamatsu {\it et al.}, Nature {\bf 410}, 63 (2001).
\bibitem{Kortus} J. Kortus {\it et al.}, Phys. Rev. Letters {\bf 86}, 
4656 (2001).
\bibitem{STM} F. Giubileo {\it et al.}, Phys. Rev. Letters {\bf 87}, 
177008 (2001).
\bibitem{MultiGap} J.W. Quilty {\it et al.}, Phys. Rev. Letters {\bf 90}, 
207006 (2003). F. Bouquet {\it et al.}, Phys. Rev. Letters {\bf 87}, 
047001 (2001). S. Souma {\it et al.}, Nature {\bf 423}, 65 (2003).
\bibitem{Suhl} H. Suhl, B.T. Matthias, and L.R. Walker, Phys. Rev. Letters 
{\bf 3}, 552 (1959).
\bibitem{Choi} H.J. Choi {\it et al.}, Nature {\bf 418}, 758 (2002).
\bibitem{Kang} W.N. Kang {\it et al.}, Science {\bf 292}, 1521 (2001) .
\bibitem{Carr} G.L. Carr {\it et al.}, Phys. Rev. Letters {\bf 85}, 
3001 (2000).
\bibitem{Lobo} R.P.S.M. Lobo {\it et al.}, Rev. Sci. Instrum. {\bf 73}, 
1 (2002).
\bibitem{Zimmermann} W. Zimmermann {\it et al.}, Physica C {\bf 183}, 
99 (1991).
\bibitem{IR} J. H. Jung {\it et al.}, Phys. Rev. B {\bf 65}, 
052413 (2002). R.A. Kaindl {\it et al.}, Phys. Rev. Letters {\bf 88}, 
027003 (2002).
\bibitem{UF1} Y. Xu {\it et al.}, Phys. Rev. Letters {\bf 91}, 
197004 (2003).
\bibitem{UF2} J. Demsar {\it et al.}, Phys. Rev. Letters {\bf 91}, 
267002 (2003).
\bibitem{Owen} C.S. Owen, and D.J. Scalapino, Phys. Rev. Lett. 
{\bf 28}, 1559 (1972).
\bibitem{Elliott} R.J. Elliott, Phys. Rev. {\bf 96} 266 (1954).
Y.Yafet, Solid State Phys. {\bf 14}, 1 (1963).
\bibitem{Monod} P. Monod, and F. Beneu, Phys. Rev. B {\bf 19}, 
911 (1974).
\bibitem{Simon}	F. Simon {\it et al.}, Phys. Rev. Letters {\bf 87}, 
047002 (2001).
\bibitem{Urbano} R.R. Urbano {\it et al.}, Phys. Rev. Letters {\bf 89},
087602 (2002).
\bibitem{Eltsev} Yu. Eltsev {\it et al.}, Phys. Rev. B {\bf 66},
 180504 (2002).
\bibitem{Shukla} A. Shukla {\it et al.}, Phys. Rev. Letters {\bf 90}, 
095506 (2003).

\end{chapthebibliography}
\end{document}